\documentclass[10pt, conference]{IEEEtran}
\IEEEoverridecommandlockouts
\usepackage{cite}
\usepackage{amsmath,amssymb,amsfonts}
\usepackage{algorithmic}
\usepackage{graphicx}
\usepackage{textcomp}
\usepackage{xcolor}
\usepackage[nounderscore]{syntax}
\usepackage{booktabs}
\usepackage[export]{adjustbox}
\usepackage{makecell, cellspace, caption, subcaption, longtable, array}
\def\BibTeX{{\rm B\kern-.05em{\sc i\kern-.025em b}\kern-.08em
    T\kern-.1667em\lower.7ex\hbox{E}\kern-.125emX}}
\begin{document}

\title{Abstractions for Network Intelligence:
A Reference Architecture for AI at the Wireless Edge
}

\author{
  \IEEEauthorblockN{
    Salil Reddy\IEEEauthorrefmark{1},
    Haohuang Wen\IEEEauthorrefmark{1},
    Ness Shroff\IEEEauthorrefmark{1},
    Venki Ramaswamy\IEEEauthorrefmark{2}\\
    Zhiqiang Lin\IEEEauthorrefmark{1},
    Elisa Bertino\IEEEauthorrefmark{3},
    Jim Kurose\IEEEauthorrefmark{4},
    Anish Arora\IEEEauthorrefmark{1}
  }
  \IEEEauthorblockA{\IEEEauthorrefmark{1}The Ohio State University}
  \IEEEauthorblockA{\IEEEauthorrefmark{2}MITRE Labs}
  \IEEEauthorblockA{\IEEEauthorrefmark{3}Purdue University}
  \IEEEauthorblockA{\IEEEauthorrefmark{4}University of Massachusetts, Amherst}
}

\maketitle

\begin{abstract}
Networks are increasingly adopting AI as are AI applications leveraging networks. Awareness sharing between networks and AI applications promises to unlock higher levels of network utilization and application performance, but is inadequately supported in the current architecture of the Internet. In this paper, we describe a reference architecture that abstractly enables the synergistic interaction of intelligent applications and the intelligent network, via an information waist, and also supports the network intelligence services in the emerging intelligence plane in networks. We discuss the rationale for our AI-EDGE architecture, its functional requirements, and the core abstractions. We present a reference component-level design of the core abstractions to support experimentation and development on existing platforms for wireless networking (i.e., based on O-RAN cellular networks) and edge computing (i.e., based on 3GPP Edge App and ETSI MEC). Moreover, we provide representative use cases from the perspective of different types of users that demonstrate the benefits of the architecture in contexts of awareness sharing, portability, prototyping, and validation. 
\end{abstract}

\vspace*{1mm}
\begin{IEEEkeywords}
network architecture, AI/ML, awareness, information waist
\end{IEEEkeywords}

\section{Rationale}
\label{sec:rationale}

The internet has remarkably continued to evolve hand in hand with emerging forms of use. Over the past two decades, it has incorporated the Internet of Things, accommodated the onslaught of Big Data, and enabled Cloud Computing of large-scale workloads.  More recently, the internet has grown notably at the edge, with deployments of new wireless last mile and middle mile technologies such as 5G. A generational shift from the centralized, core Cloud to a distributed, wireless Edge is ongoing.

\begin{figure*}[h]
\begin{center}
\includegraphics[width=.8\textwidth]{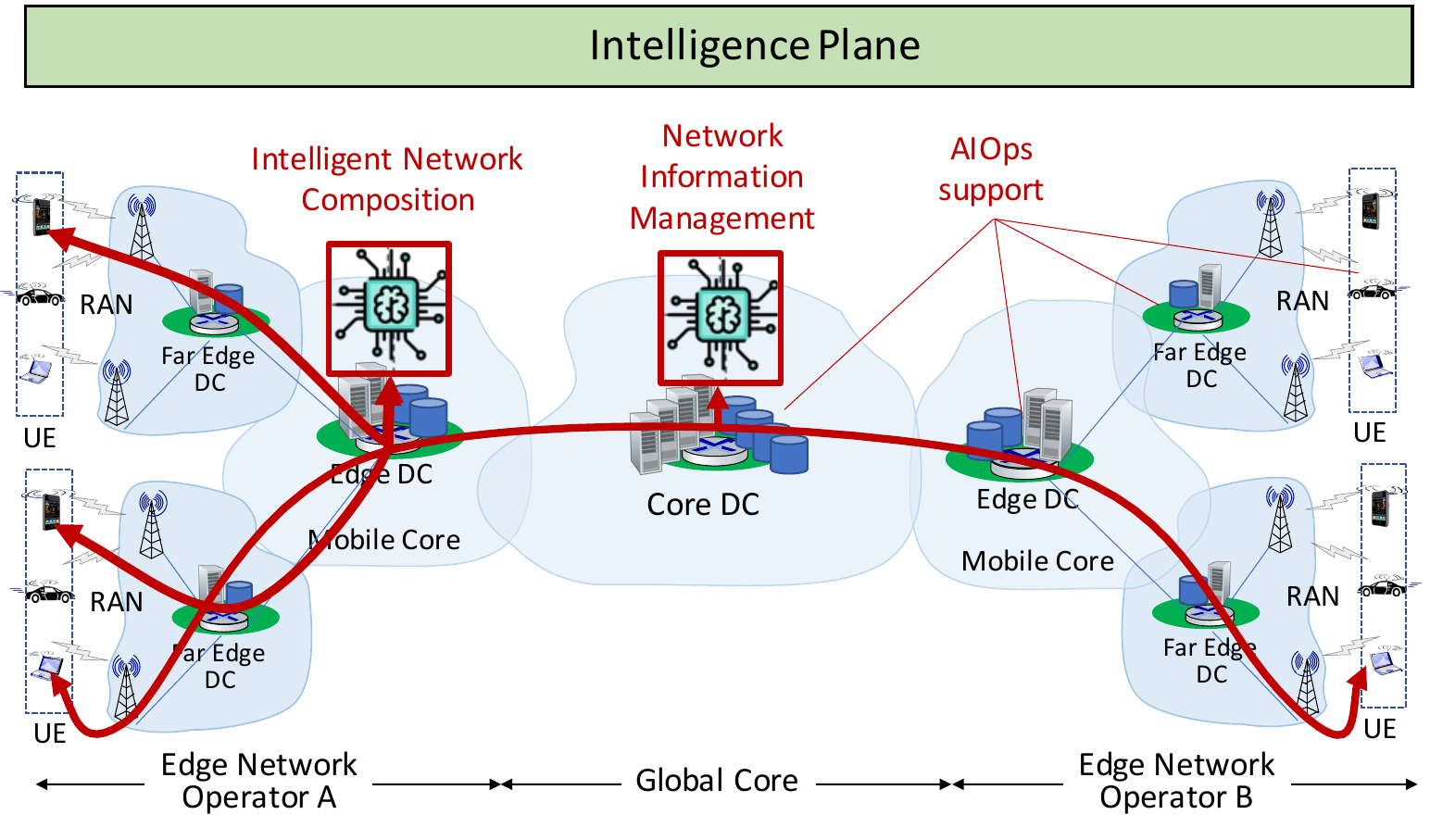}
\end{center}
\caption{End-to-end view of network intelligence operations of the intelligence plane in a multi-operator wireless edge network. (DC refers to data center). The AI-EDGE reference architecture comprises core abstractions to support sharing for AI-on-networks and AI-for-networks and these operations.} 
\label{fig:AI-EDGE}
\end{figure*}

A new {\em Intelligence Plane} is emerging for the Internet, as its workload is increasingly shifting towards Artificial Intelligence (AI) and Machine Learning (ML) applications and as the network increasingly adopts AI/ML models for its own operation \cite{letaief2019roadmap}.  For the former category, aka AI {\em on} Networks, the Intelligence Plane supports AI applications, by sharing network awareness, in reaction to which  application or their managers can intelligently act or adapt. And for the latter category, aka AI {\em for} Networks, the intelligence plane optimizes network utilization, control, and  management, using relevant data, models, and tools. (See {\bf Figure} \ref{fig:AI-EDGE}.)

There is by now substantial research and development on ---and even bespoke solutions implemented by operators of---  network support and incorporation of AI/ML models. In the context of 5G cellular networks, the outcome of explorations in 5G + AI has been defined by some as the 6G network. Nevertheless, the intelligence plane is not explicitly reflected in today's internet architecture. There is a lack of standards, which understandably take time to develop, but more importantly there is also a lack of a reference architecture of how intelligence is natively supported and integrated in networks, especially those at the wireless edge.

\vspace*{1mm}
\noindent
{\bf Motivation for our Reference Architecture}.~~~Significant gaps exist in the current network architecture:  Network awareness sharing with applications appears in rather limited forms. The converse---application awareness sharing with networks--- is with little exception missing. The reference architecture thus serves to incorporate network abstractions that support rich but trustworthy sharing of awareness between applications and the networks. It also serves to include other core abstractions that support the broader functionality of the intelligence plane (cf.~Figure~\ref{fig:AI-EDGE}).  Moreover, as corroborated by our experience working with a diverse set of stakeholders actively engaged in research, design, prototyping, and product development of both intelligent edge networked systems and applications, it serves important needs from each of their perspectives, which we outline below.

For researchers, mapping their AI solutions to components of the architecture leads to a better understanding of the designs and helps in validating the assumptions made. For example, the mapping clarifies the use of various planes---such as the user, control, management, and intelligence plane---in their designs as well as the locations of different components with respect to networks/devices.  In turn, this reveals the resource and access constraints of the network elements being used, which may arise from infrastructural limitations or organizational policies. These clarify the feasibility of the AI design, especially in cases where the AI runs on embedded edge devices or across network operator boundaries.

With respect to AI designers, the architecture helps in crossing a large chasm for validating the effectiveness of their algorithms for realistic applications. Several unknowns factor into how their algorithms will work in practice: They may not have access to the applications themselves. They may even lack access to the user platform or infrastructure on which the applications operate. Framing their design in terms of the central concepts (i.e., a waist or a core API) and services of the reference architecture can make it easier to achieve a first cut for simulation, prototyping, or validation.

From the perspective of AI developers, the abstractions provided by a reference architecture help to mask platform-specific complexities from their task. In addition, reference architectures typically have associated implementations in one or more reference platforms. These reference implementations provide yet another stepping stone for AI developers towards realization of fielded networks and applications. For example, reference implementations in virtualized platforms enable fast prototyping and convenient testing.  Tools incorporated in reference implementations further simplify the engineering of innovative solutions. Also, reuse of reference implementations reduces the development time and effort in porting from one platform to another.

Reference architectures tend to be open---even if only in the form of openly published APIs. This openness not only facilitates experimentation by diverse stakeholders, it also help network operators avoid lock-in with respect to proprietary or vendor-specific implementations of key modules. The freedom to plug-and-play with different modules can also facilitate interoperability of networks.

Lastly, reference architectures serve to inform the process of standardization. By the same token, they can help in future proofing for upcoming generations: The architecture can be ported to standalone implementation for a future generation network; alternatively, a reference implementation of a current generation can be extended to support an upcoming generation.  Just as the 5G Non-Standalone implementation has been built upon existing 4G networks, the realization could yield a platform that supports both a current generation and an upcoming generation, allowing for backward compatibility. It is our hope that our reference architecture thus contributes to the standardization of 6G. 

\vspace*{1mm}
\noindent
{\bf Outline and Contributions of the Paper}.~~~We present a reference architecture that incorporates an intelligence plane for a network to adequately support and leverage AI/ML. Since the architecture emphasizes considerations at the wireless edge of the network, we call it the {\em AI-EDGE} architecture. 

The core abstraction in AI-EDGE for achieving awareness sharing is an {\em Information Waist}. The waist allows applications and network entities to specify sharing needs abstractly, as information. Computation of the information in-network is orchestrated by the waist, with minimal engagement, to securely and efficiently establish the satisfiability of specifications and execute relevant components for the computation.  We show that the waist suffices to embody rich patterns of interaction for AI. 

Other core abstractions ---{\em Reactivity Engine, Intelligence Orchestrator}, and {\em Policy Engine}--- provide support for common services of the intelligence plane (cf.~Figure~\ref{fig:AI-EDGE}). We detail the design of AI-EDGE abstractions, which reuse networking advances that have gained traction and validation in clean slate architectures over the past two decades, i.e., named data and named functions. The design is minimal and focuses on extensibility, openness, and scalability of intelligence plane services.

Lastly, we validate the AI-EDGE architecture through several use cases, each illustrating different motivations for the reference architecture discussed above. In particular, we show that it suffices to embody rich patterns of interaction in AI-on-networks and AI-for-networks. These use cases have been developed by our partner stakeholders. We note that stakeholder engagement has also guided the framing of its requirements as well as scoping of its network abstractions.

\section{Requirements for AI-EDGE Reference Architecture}
\label{sec:requirements}

Network control increasingly requires a deep understanding of the network and the ability to make network (and other) information rapidly accessible across layers and across planes. The AI-EDGE architecture incorporates this understanding and sharing
in a singular dimension which exists as its own plane and permeates the rest. This so-called intelligence plane is solely responsible for enabling knowledge representation, gaining and imparting information of the network, and realizing information-driven network control, without violating separation of concerns. 

The idea of a dedicated plane that encodes and shares information awareness for control is hardly new; a precursor has been proposed in the form of a ``knowledge plane'' \cite{clark2003knowledge, bullot2008situatedness, manoj2008cognet, camelo2022daemon}. That conceptualization predicted the use of ``cognitive'' techniques in lieu of traditional analytic solutions as the basis for knowledge. Two decades later, a knowledge plane has yet to be adopted \cite{mestres2017knowledge, ashtari2022knowledge}, but given early successes in the adoption of AI-for-networks, we have the benefit of building on remarkable advancements in AI/ML since then and can embrace AI/ML models as the units of intelligence that encode and act on information, if not on knowledge. 

It is important to distinguish the intelligence plane, which leverages information-based awareness sharing as a primitive, from Information-Centric Networking (ICN) \cite{ghodsi2011information, ahlgren2012survey}.  ICN focuses on communication and, in some versions, on in-network computation \cite{sifalakis2014information, krol2019compute} of information; it has been demonstrated for network optimization functions \cite{mastorakis2020icedge} and AI/ML applications \cite{mtibaa2018towards}. However, from an architectural perspective, it does not focus on the intelligence plane or on refactoring network planes; nor has it been used as a basis for awareness sharing. 

\subsection{Awareness Sharing}

A primary requirement of the AI-EDGE architecture is to incorporate functions in the Intelligence Plane that support AI on Networks and AI for Networks.  
For supporting the former, AI applications need a way to specify and obtain desired ``network awareness'', given which they can configure or adapt themselves. Conversely, they also need a way of specifying appropriate ``application awareness'' with which the network can provision or maintain resources to support the AI applications effectively. For supporting the latter, 
i.e., supporting intelligent optimization of network control, management, and the like, the intelligence plane itself needs access to network awareness corresponding to targeted network components.

Note that this requirement of sharing application- and network-awareness entails transfer of information across layers---from network to application and/or vice versa--- and across planes---say from the data, control, and management plane to the intelligence plane.  In this sense, the intelligence plane needs to act as a broker for bidirectional propagation of information. We frame the sharing of awareness in terms of information as opposed to data: Information is more abstract than data, and it efficiently captures what needs to be shared while offering degrees of freedom in the ``what'' and ``how'' aspects of the sharing, which are well suited to achieving a high-performance network that AI-on-networks today demands.
Nevertheless, the sharing of information must be designed to be easy to use, to avoid creating undue overhead or complexity, and importantly to preserve security and privacy. 

\subsection{Support for Network Intelligence}

The intelligence plane is likely to vary from operator to operator and even from network instance to instance. It is also likely to expand in scope given the rapid pace of advances in AI in general and in AI for networking. Therefore, the reference architecture intentionally eschews offering of all intelligence plane services.  Instead, its requirement is to offer core abstractions that support commonly used intelligence plane services. 

Towards identifying candidate core abstractions, let us briefly review the scope and considerations of common intelligence plane services:
\begin{itemize}    
    \item {\bf Network Composition}:
    As an example, intelligent network optimization entails slicing, a network operation that allocates available resources to network functions and application traffics. In addition, it entails dynamic reconfiguration of the network in response to changes in usage and the network environment. 
    
    \vspace*{1mm}
    \item {\bf Network Information Management and AIOps}:
    The intelligence plane typically manages network data and ML models for its network AI operations.  It supports interfaces for discovery of these data and models, locally as well as from external sources and vendors. These data and models may also be accessed and updated by network modules and functions residing in other planes. In addition, it provides services for orchestration of network AI operations. 
    
    \vspace*{1mm}
    \item {\bf Network Security and Privacy}:
    The compositional nature of networks and in turn the intelligence plane imply that they often span across boundaries of not just operators, but also users and intents. Trust between networks spanning such boundaries and even network components is tenuous, and entails enforcement of corporate policies and liability considerations that constrain sharing of information, which may include confidential and proprietary data. These services facilitate gatekeeping of network information across a multitude of entities within and outside the network while allowing for integration with policy engines that can dynamically control policies regarding access and authorization.
\end{itemize}

As expected, several of these services can themselves use the abstractions for awareness sharing and rely on AI/ML. More importantly, they can share core abstractions, notably, that provide reactivity, orchestration of intelligence operations, and policy enforcement in a generic fashion. Beyond supporting multiple users, core abstractions in the reference architecture should meet various desiderata, such as:
\begin{itemize}
\item {\bf Minimality}:
They should avoid as far as possible the prescription of specific mechanisms, and emphasize minimal specification of interfaces of the functions. By doing so, diverse implementations on potentially heterogeneous platforms and interoperability can be facilitated. Backwards compatibility with respect to legacy protocols and systems is also facilitated. Specific implementations can choose their own mechanisms.  

\item {\bf Security and Privacy}:
Abstractions potentially expose a range of new attack vectors in the intelligence plane. Adversaries may include malicious network nodes and applications that could eavesdrop on, intercept, spoof sensitive traffic data, or flood the network with poisoned data. Likewise, there may be intrusions, malware activity, and internal threats such as advanced persistent threats (APTs). These threats must be addressed. 

\vspace*{1mm}
\item { \bf Extensibility}:
Implementers and clients of abstractions should be able to extend their interfaces or functionality.
\end{itemize}

\section{Architectural Abstractions in AI-EDGE}
\label{sec:elements}
 
\begin{figure}[b]
\begin{center}
\includegraphics[width=0.485\textwidth]{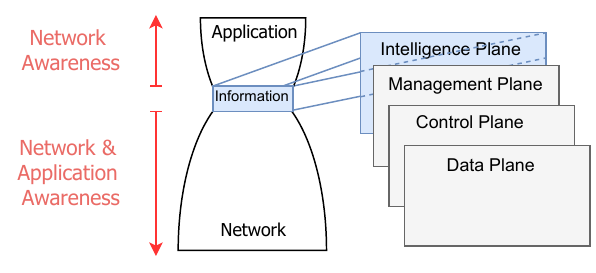}
\end{center}
\vspace*{-2mm}
\caption{The Information Waist supports the functionality of the Intelligence Plane for disseminating network awareness to applications and both network and application awareness within the network. The support may itself be realized by exploiting AI/ML.} 
\label{fig:Info-waist}
\vspace*{-4mm}
\end{figure}
\noindent

\vspace*{-2mm}
\subsection{The Information Waist}
\label{sec:realIP}
The information waist is a layer between AI applications and the network. It is a waist in the sense that it offers a narrow, elemental abstraction for structuring the interactions between higher and lower layers. Just like the Internet Protocol (IP) waist, introducing the information waist also implies that other interactions between the higher and lower layers are eschewed.  See Figure~\ref{fig:Info-waist}, which also clarifies that the waist is part of the Intelligence Plane. As such, using the waist allows for intelligent management of the network itself, as well as of applications.

\vspace*{3mm}
\noindent
{\bf Specifying Information}.~~~
To specify information minimally, we reuse the well-known concepts of named data \cite{jacobson2009networking, ghodsi2011naming, zhang2014named}, and named functions \cite{tschudin2014named, sifalakis2014information, mtibaa2018towards}. Syntactically, information $I$ can be specified as the tuple $(D,F)$, where $D$ is a named datum and $F$ is a named function. Semantically, $I$ evaluates to $F(D)$. 

In the degenerate case where the named function is the identity (or no-op) function,  
the information specification reduces to that in Named Data Networking (NDN) and similar proposals for Information-Centric Networking (ICN). More generally, we allow information to be specified compositionally over vectored data and vectors/sequences of functions, allowing for sequential or parallel evaluation as desired. 
Sequential specifications of information involve chaining of functions.  For example, the information specification$( D, F_1, F_2, ... , F_n)$, where $F_1..F_n$ are named functions, evaluates to $F_n(...(F_2(F_1(D))))$. Likewise, parallel specifications of information involve vectors of named data and/or named functions, whereby per-element evaluations yield vectorized information. Illustrative examples include: 

\begin{itemize}
\item 
$(D, [F_1, F_2)])$ evaluates to $[F_1(D), F_2(D)]$.
\vspace*{1mm}
\item 
$([D_1, D_2], F)$ evaluates to $[F(D_1), F(D_2)]$, for named data $D_1, D_2$. 
\vspace*{1mm}
\item 
$([D_1, D_2], F_1, F_2)])$ evaluates to $[F_2(F_1(D_1)), $ $F_2(F_1(D_2))]$. 
\vspace*{1mm}
\item 
$([D_1, D_2], [F_1, F_2])$ evaluates to
$\left[    {\begin{array}{cc}
   F_1(D_1) & F_1(D_2)
   \\
   F_2(D_1) & F_2(D_2)
   \\
 \end{array} } \right]
$. 
\end{itemize}

\vspace*{1mm}
For cases where named functions have multiple arguments, information specifications include a matching set of named data. Thus, for example, given a dyadic function $F$, the information tuple $((D_1, D_2), F)$ evaluates to $F(D_1, D_2)$. 

Information specifications allow information to itself be treated as named data, thereby allowing for composing of information from other information. Similarly, they allow named functions to be considered as a form of named data, thereby allowing, for example, an AI/ML model to be used not only as a named function but also as named data. Similarly, some forms of information $(D, F)$ may be used as named functions provided $F(D)$ itself evaluates to a function. Figure \ref{fig:grammar} expresses the grammar of information specifications over literals of named data $ND$ and functions $NF$.
\vspace*{1mm}
\begin{center}
\fbox{\begin{minipage}{7cm}
\begin{grammar}
<information> ::= `(' $\langle$data$\rangle$ `,' $\langle$func$\rangle$ `)'  
\vspace*{1mm} \alt `(' $\langle$data$\rangle$ `)' 

<data> ::= $ND^+$  
\vspace*{1mm} \alt `[' $ND^+$ `]'
\vspace*{1mm} \alt $\langle$$in\!f\!ormation$$\rangle$

<func> ::= $NF^+$
\vspace*{1mm} \alt `[' $NF^+$ `]'
\end{grammar}
\captionof{figure}{Grammar for specifying information}
\label{fig:grammar}
\end{minipage}}
\end{center}

\vspace*{2mm}
\noindent
{\bf Communicating Information}.~~~Sharing of information by the waist is based on the primitive patterns of publish-subscribe.  Named objects must be published by either applications or the network before they can be located and utilized. The \emph{PUBLISH} primitive is used to declare both data or function type named objects. The \emph{SUBSCRIBE} primitive creates a demand in the information waist for the corresponding information, whose named object components are located, routed, and appropriately evaluated and the resultant evaluation delivered to the subscriber of the information request.

This formulation of the information waist treats data and functions as primitive network entities and lends itself to leveraging existing work on named data and named functions in networking. The choice of pub-sub also inherits from the prevalent communication mechanism in NDN and ICN architectures, as well as in software-defined networking (SDN) platforms \cite{ferguson2021orion}. However, while previous work on NFN and named functions proposed a lambda calculus for specifications \cite{tschudin2014named}, the minimal grammar in Fig.~\ref{fig:grammar} is sufficiently expressive to support the functionality of the intelligence plane as envisioned. 

\vspace*{2mm}
\noindent
{\bf Computing Information In-Network}.~~~While information specifications, publishers, and subscribers capture the ``what'' aspect of evaluating and sharing information, the ``how'' aspect is left up to platform-specific implementations of the information waist. Each waist implementation is responsible for selecting its mechanisms, protocols, and policies for publishing, subscribing, and coordinating the evaluation and delivery of information. These selections would resolve, for instance, how to locate named objects, choose appropriate locations for sourcing objects in cases where copies of the objects are available at multiple locations, migrate as need be either named data or named functions (if not both) to the location for evaluation of information, invoke the evaluations, and choose appropriate protocols to transport the resulting information to their subscribers. A simple implementation may rely on pre-defined policies for each of these decisions, whereas a more intelligent one would leverage support functions of the intelligence plane to optimize the decisions.

A remark is in order for how the waist would coordinate evaluation of information. Named functions are evaluated by their respective owners, in their respective networking or application domains. This approach helps to not only contain the scope of the information waist, it also avoids trust issues in the information waist associated with named objects. That said, allowances or constraints specified by the owners regarding named functions or data are to be followed by the information waist to coordinate their evaluation. These are incorporated in the form of metadata associated with the named objects.  For example, metadata associated with a named data at an edge location in the network may specify that the data is ``pinned'', i.e., proscribed from being transported beyond the edge, say, for reasons of privacy or to limit its users for reasons of confidentiality. In this case, the coordination by waist would only evaluate allowed functions on the data at the edge, which would involve transporting the functions from their source location to the edge location rather than the other way around. 
Similarly, a named data may specify supported transport protocols as part of its metadata whereas a named function's metadata could specify support for a particular micro-architecture. Evaluating information must adhere to these constraints.

\subsection{Core Abstractions for Supporting Network Intelligence}

In addition to the information waist, the AI-EDGE architecture provides the following core abstractions to support the common intelligence plane services described in Section~\ref{sec:requirements}:

    \vspace*{2mm}
    \noindent
    {\bf Reactivity Engine}.
    This abstraction serves intelligence plane services (and other intelligent network services) that need to adapt to the dynamic nature of the network or deal with occurrences of special network conditions, by activating appropriate actions in response. Users of the engine need to specify the changes to be detected and the actions to be performed. In turn, the engine will use the information waist to obtain the network awareness of the changes at hand, instantiate change detection, and instrument logical rules or AI models associated with the actions at hand. 
    
    \vspace*{2mm}
    \noindent
    {\bf Intelligence Orchestrator}.
    This abstraction provides core support for AIOps and model life cycle management, which are increasingly important as network intelligence proliferates. The orchestrator services network AIOps by providing simple interfaces for locating, provisioning, and updating intelligent models while leveraging the underlying information waist where these models exist as named data or functions. Furthermore, the orchestrator can use the reactivity engine to facilitate generic closed-loop automation in the intelligence plane. For example, the reactivity engine can trigger model updates in the orchestrator itself based on changing network conditions. 

    \vspace*{2mm}
    \noindent
    {\bf Policy Engine}.
    This abstraction supports security, privacy, and cross-operator services in the intelligence plane. Services can use the engine to assign appropriate access policies associated with data and models in the network or to re-evaluate model policies based on changing network state and ownership.  Network operators can use the engine for restricting cross-operator information sharing. The engine moreover includes options to verify and correct improper policy assignments, which can undermine security and privacy. 

\section{Reference Architecture Design}
\label{sec:impl}

In this section, we present a reference design for AI-EDGE abstractions, in terms of their generic components, that meets the desiderata for abstractions prescribed in Section~\ref{sec:requirements}.

\subsection{Information Waist Design}

\begin{figure}[b]
\begin{center}
\includegraphics[width=0.475\textwidth]{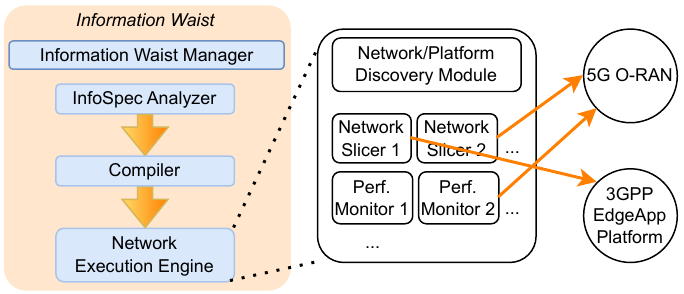}
\end{center}
\caption{Component-level design of Information Waist.}
\label{fig:ref_design}
\end{figure}

We compose the waist abstraction as follows (see Figure~{\ref{fig:ref_design}}):
\begin{itemize}
    \item {\bf Information Waist Manager}:
    This component monitors incoming information requests and performs admission control, scheduling, and bookkeeping for the information waist.
    
    \item {\bf InfoSpec Analyzer}:
    This component parses and type-checks information specifications. It also processes necessary metadata of each named data and function for compilation purposes.
    
    \item {\bf InfoSpec Compiler}:
    This component generates code that describes provisioning, scheduling, and communication in the network needed to satisfy a given information request. The network directives output by the compiler are agnostic with respect to network implementation, i.e., named functions are yet to be mapped to their corresponding concrete implementations of the underlying network.
    
    \item {\bf Network Execution Engine}:
    This component executes the network code generated by the compiler. To do so, it needs to map the named functions provided by the compiler to their concrete implementation in the network where the function is to be executed.
    It uses a network discovery service that determines the type of underlying network implementation and ensures that appropriate functions are invoked in the underlying network for resource provisioning, slicing, orchestration, network monitoring, etc.
    In addition to scheduling network function execution, the engine sets up the requisite communication flows between the translated endpoints of the named objects. Clients can choose whether the execution will  compute the requested information (in cases where information is generated within reasonable time scales) or a handle at which to receive that information (in cases of long-running subscriptions).
\end{itemize}

\subsection{Design of Core Abstractions}

Our designs of the core abstractions include components that interact with one another as well as with the underlying information waist. 

\vspace*{1mm}
\noindent
{\bf Reactivity Engine}. Recall that the engine interface allows clients to specify a set of network conditions to be observed and a corresponding set of actions to be executed in response.  Two components suffice for its design:
        
        \begin{figure}[b]
        \begin{center}
        \includegraphics[width=0.475\textwidth]{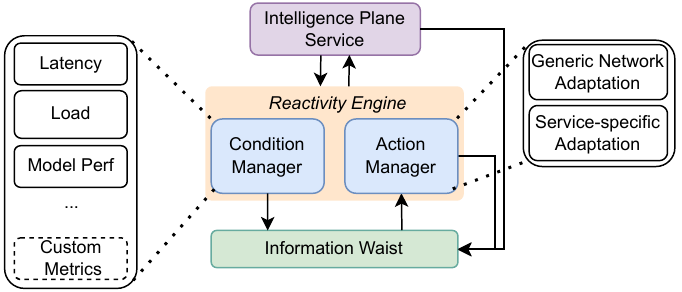}
        \end{center}
        \caption{Component-level design of Reactivity Engine.}
        \label{fig:ref_react}
        \end{figure}

 \begin{itemize}
    \item   {\bf Condition Manager}: 
            This component maps abstract conditions into information derived from concrete network metrics. It allows clients to substitute its mapping logic with custom metrics (this is similar to functionality provided by the E2 Service Model (E2SM) \cite{alliance2020ran} in the O-RAN RIC architecture \cite{o2021ran}). Finally, it invokes the information waist to obtain the information about concrete conditions. The information can be of various granularities and obtained in batch or stream mode.
            
            \item {\bf Action Manager}:
            This component matches network conditions to their response actions. It provides default network adaptation responses to common network conditions, such as network resource load balancing, but also allows clients to extend the response such that custom actions (based on, say, reinforcement learning models) are incorporated into the engine.
\end{itemize}

\vspace*{1mm}
\noindent
{\bf Intelligence Orchestrator}.  For this abstraction to support network AIOps and other network intelligence services that require native network support for model management, its interfaces allow clients to specify their models of interest and a corresponding set of life-cycle management operations. It has the following components (see Figure~\ref{fig:ref_intello}):
        
        \begin{figure}[b]
        \begin{center}
        \includegraphics[width=0.3\textwidth]{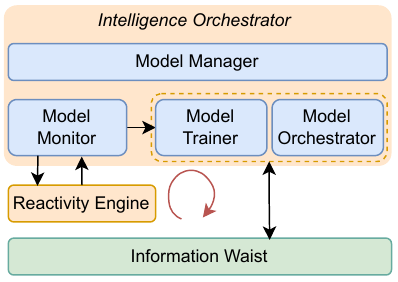}
        \end{center}
        \caption{Component-level design of Intelligence Orchestrator.}
        \label{fig:ref_intello}
        \end{figure}

\begin{itemize}
        \item {\bf Model Manager}:
            This component manages the various models specified by clients and generates management plans for each model based on the associated life-cycle requirements. For example, in an O-RAN 5G network, the objective of an intelligence plane service could be to optimize network reliability in a particular region. The model manager in this case would have to do the following: identify the associated intelligent models ---xApps, rApps, etc.--- across multiple planes involved in the optimization; deploy or replace the models at chosen locations, e.g., at GNodeBs in the corresponding network coverage region; and invoke model performance monitoring and conditional model re-training for some or all models as per the service's specifications.
            
            \item {\bf Model Orchestrator}:
            This component uses the information waist to orchestrate models as named functions. 
            
            \item {\bf Model Trainer}:
            Given a model and its data dependencies, this component uses the information waist to set up appropriate data flows for the model to be re-trained on new training data. Invocation of the information waist will involve a training script that is a named function, and the model along with its training dataset that are named data. The exact training script can either be provided as the service or be associated with the specific model such that it can be queried from an external source.
            
            \item {\bf Model Monitor}:
            This component monitors model performance by configuring the reactivity engine to invoke other components of the intelligence orchestrator in response to its monitoring. For example, the response could be to replace a model with another, in which case the model orchestrator would be called to deploy another model. Or, degrading performance could trigger model re-training by invoking the model trainer. Thus, the model monitor can also incorporate support for closed-loop model management.
\end{itemize}
    
\vspace*{1mm}
\noindent
{\bf Policy Engine}. Security and privacy-related services in the intelligence plane typically need to configure and maintain access control policies for (model, data, and service) entities that are accessible through the plane. Moreover, they require access to security-related information of all intelligence plane operations. Sharing between intelligence planes of different operators also needs to be governed by policies. This engine supports such functionality with a generic {\bf Access Control Policy Manager} component. The component generates and manages access control policies (for authentication and authorization) of each named data/function and the entities in the intelligence plane. It allows clients to selectively query the manager for policies of services and users in the intelligence plane.

The design does not restrict users from create policies in different ways, say, using membership and association rules or by learning from historical access request records \cite{katsis2025zt}. Moreover, its support of querying for policy information caters to active auditing and rapid security incident response, enabling practical and rigorous security compliance. 

\subsection{Conformance with Desiderata}

{\bf Minimality}. 
Occam's razor is at play in all AI-EDGE abstractions. Services for discovery, transport, and computation associated with named objects, etc., are relegated to being outside of the waist and other core abstractions. Most choices of mechanism and policy for the abstractions are left up to the implementer. The information waist, for instance, is intentionally limited to a minimal set of concepts. Implementation of its central concepts, such as named data and functions, leverage significant previous experience and can reuse existing implementations of  NDN \cite{thompson2014ndn, newberry2021yanfd} and NFN \cite{scherb2019execution, kumamoto2020real}. Similarly, the Policy Engine implementations can borrow from existing access control schemes for NDN \cite{nour2021access} while also utilizing the underlying standardized access control implemented of the 5G O-RAN or the ETSI Multi-access Edge Compute (MEC) \cite{kekki2018mec} and 3GPP EdgeApp \cite{3gppedgeapp} platforms. 

How intelligent the abstractions will themselves be is left as an implementation and platform-dependent consideration. That said, AI/ML has found significant use in network composition, resource provisioning and management, learning policies, and even for network security and privacy.  For instance, several application aware approaches to intelligent network slicing have been developed in recent years to produce rightsized network slices per application (i.e., \cite{righteous, zhang2021sinan, shu2020novel}). Network operators have demonstrated spectrum efficiency and energy savings gains by intelligent profiling of network traffic and corresponding control \cite{d2022orchestran}. We thus expect a significant role for AI/ML in implementations leveraging the AI-EDGE architecture.

\vspace*{1mm}
\noindent
{\bf Security and Privacy}.
The information waist and core abstractions potentially expose a range of new attack vectors in the intelligence plane. Our threat model considers both external adversaries and internal threats. Potential adversaries may include compromised network nodes and malicious applications that could: seek unauthorized data access; eavesdrop on, intercept, or spoof sensitive traffic data; flood the network with poisoned data; or disrupt service availability. 

To address these threats, our design recommends a {\it Zero Trust} implementation approach to ensure both security and privacy, and that corresponding zero-trust-based policies and mechanisms be enforced. In addition to recommending access control measures to restrict data and resource access, the design recommends that implementations leverage encryption and digital signatures for data and information transmission processes to ensure confidentiality and integrity. Network functions and data should be properly registered and verified by trusted mechanisms before being published, particularly for privileged functions that require kernel-level operations. Advanced detection and analytics algorithms should also be employed to identify intrusions, malware activity, and internal threats such as advanced persistent threats (APTs). 

Trust issues associated with the information waist merit special attention. The architecture captures named objects of two distinct types ---application and network--- whose mingling allows the waist to encode awareness of one another. This decision is not without cost. For performing in-network computing that involves objects defined by applications, the information waist must take necessary precautions that assume minimal trust of applications, while at the same time satisfying their privacy and confidentiality requirements. (Recall from Section~\ref{sec:elements}, in-network computation of information involves cooperation at evaluation time between the respective domains that published the named objects being evaluated, while also respecting their allowances and enforcing their constraints.) The same applies for AI-EDGE computations across potentially untrusted (operator) network boundaries, introducing security concerns unique to distributed, cooperative AI systems. Trust is modeled dynamically, assuming network nodes under a single operator have high trust, while cross-operator interactions and all sharing between applications and networks is subject to rigorous trust verification.

\vspace*{1mm}
\noindent
{\bf Extensibility}.
The design explicitly supports extensibility of the abstractions: In the information waist, this is be achieved by using metadata for named data and functions. And in the other core abstractions, clients can extensibly specify inputs, such as conditions, actions, life-cycle operations, and policies, and functions, such as reactivity mappings.

\section{Architectural Use Cases}
\label{sec:usecase}

We now showcase select use cases from our partner stakeholders' research and development based on the use of the AI-EDGE Intelligence Plane architecture. The use cases are representative of common patterns of AI-on-networks and AI-for-networks in edge networking, demonstrate different features of AI-EDGE, and capture different perspectives: the user is, respectively, an AI researcher, an AI application manager, and an AI systems developer in the three use cases.

\subsection{Data Selection in Federated Learning}

\begin{figure}[b!]
\includegraphics[width=0.45\textwidth]{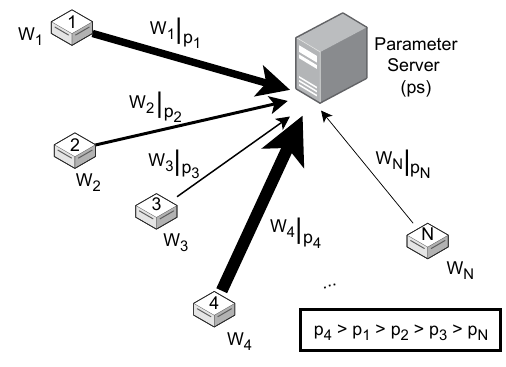}
\vspace*{-3.5mm}
\begin{equation*}
I_{ps} = ([W_{1} \, .. \, W_{N}], ((([(W_1, ps) \, .. \, (W_N, ps)], F_{bw})), F_{f\!p})) 
\end{equation*}
\vspace*{-3.5mm}
\caption{Network aware federated learning weight updates.} 
\label{fig:netaware_fl}
\end{figure}

As Deep Neural Networks (DNNs) scale in data size, model size, and compute complexity, both their training and inference are typically achieved through distribution. We first consider a learning context, one where a Federated Learning application over a large heterogeneous network needs to be network aware to limit in some fair fashion the sharing of model weights depending upon the available network capacity \cite{yu2021jointly, qu2022context} (and, for ease of exposition, not depending upon other factors). More specifically, client endpoints $i, i \in \{1..N\}$ each have local model weights, $W_i$, with which a parameter server $ps$ needs to update a global model. For each $i$, only a fraction $p_i$ of the weights $W_i$ is to be communicated from endpoint $i$ to $ps$, and the federated learning application at the server calculates this fraction based on the available network capacity between all endpoints and the server.

The information $I_{ps}$ specified in Fig.~\ref{fig:netaware_fl} illustrates a way of simplifying the application's task, by delegating the coordinated acquisition of the fractional weights by the server from each endpoint by subscribing to $I_{ps}$ that is computed in-network. Assume the network publishes a monitoring function $F_{bw}$ that returns the available network capacity from a node (at which some named data exists) to another node.
Assume also that the application publishes a named triadic function $F_{f\!p}$ that, given a server node, an array of client nodes, and the corresponding array of network capacities between client nodes and the server nodes, computes a function that for each client $i$ yields a fraction, $p(i)$, denoting its ``fair proportion''. Lastly, assume that each client $i$ publishes its $W_i$ as eponymous named data. 

By subscribing to $I_{ps}$ from the waist, 
the server directly receives only the fair proportion fraction $p_i$ of the respective node weights $W_i$.
To see this, let's evaluate $I_{ps}$:
\begin{align*}
I_{ps}
&= ([W_{1} \, .. \, W_{N}], ((([(W_1, ps) \, .. \, (W_N, ps)], F_{bw})), F_{f\!p})) 
\vspace*{1mm} \\ 
&= \hspace*{4mm} {\rm \{~evaluating~} F_{bw} {\rm ~on~} [(W_1, ps) \, .. \, (W_N, ps)] ~\}
\vspace*{1mm}\\
&\;\;\;\;\; ([W_{1} \, .. \, W_{N}], (([F_{bw}(W_1, ps) \, .. \,  F_{bw}(W_N, ps)]), F_{f\!p})) 
\vspace*{1mm} \\
&= \hspace*{4mm}  {\rm \{~evaluating~} F_{f\!p} {\rm ~on~} [F_{bw}(W_1, ps) \, .. \, F_{bw}(W_N, ps)]  
\\
&\;\;\;\;\;\; \hspace*{4mm} {\rm ~yields~an~intermediate~monadic~function,~} F_{ws} \, \}
\vspace*{1mm}\\
&\;\;\;\;\; ([W_{1} \, .. \, W_{N}], F_{ws}) 
\vspace*{1mm}\\
&= \hspace*{4mm} {\rm \{~evaluating~} F_{ws} {\rm ~on~} [W_{1} \, .. \, W_{N}] {\rm ~resp. ~yields ~for} 
\\
&\hspace*{4mm} \;\;\;\;\;\; {\rm ~each~} i {\rm ~a~fraction~} p_i {\rm ~of~} W_i, \; \; W_i|_{p_i} := F_{ws}(W_{i})
~\}\\
&\;\;\;\;\;  (W_1|_{p_1} \, .. \, W_N|_{p_N})
\end{align*}

$I_{ps}$ uses the data and functions (of both the network and application) and captures intermediate information computations and their dependencies while allowing the network the freedom of ordering and placing them. However, as stated, for computing $F_{ws}(W_i)$, the network may either bring each $W_i$ to the node where $F_{ws}$ is located or move $F_{ws}$ to node $i$ where $W_i$ is located.  The former fails to capture the intention that network capacity limitations are the reason for the federated learning application is computing a fair proportion of the $W_i$ weights in the first place.  The former option can be precluded by ``pinning'' the named data to its location. Pinning is encoded in the metadata, and the information waist will thus know that its only option is to bring $F_{ws}$ to node $i$ for this computation.

Let us now consider a variation in the federated learning application, wherein it conversely needs to request the network to provision resources such that well ranked clients are capable of providing their entire $W_i$ weights \cite{deng2021fair} by virtue of being afforded suitably high bandwidth paths to the server. Accordingly, the application computes for each node $i$ a rank $r_i$ which in turn is used to specify the capacity constraint for corresponding evaluations of $F_{Slicer}$, a network slicing function. The resulting slicing produces rank-prioritized slices between the issuer, $ps$, and each node $i$.

This requirement can be expressed as an information specification, as follows.
\begin{equation*}
    I_{slices} = ([W_1, W_2, ..., W_N], F_{Slicer}) 
\end{equation*}

Subsequent subscriptions to weights from these nodes are now assured to utilize network slices right-sized by the priority associated with the edge node's weights. 

The two variations in this use case not only exemplify how an application can use information specifications to delegate nontrivial obligations to the network, they also illustrate how the application can leverage the bi-directional awareness sharing enabled by the AI-EDGE architecture. 

\subsection{Model Placement for Distributed Inference}

\begin{figure}
\centering
\begin{subfigure}[t]{0.45\textwidth}
  \includegraphics[width=0.63\linewidth,  center]{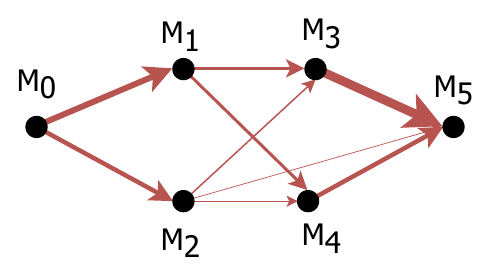}
  \caption{Directed, weighted graph where nodes are DNN model partitions, edge directions depict partition evaluation dependencies, and edge weights correspond to communication costs between partitions.}
  \label{fig:sfig1}
\end{subfigure}
\\[1ex]
\begin{subfigure}[t]{0.22\textwidth}
  \includegraphics[width=0.8\linewidth,  center]{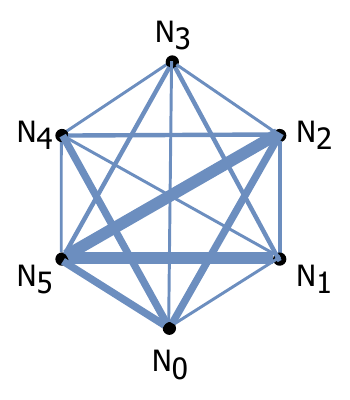}
  \caption{Network slice graph with \\edge thickness denoting \\ inter-node capacity, used \\ by network orchestrator \\ for evaluating slice fitness.}
  \label{fig:sfig3}
\end{subfigure}
\begin{subfigure}[t]{0.24\textwidth}
    \includegraphics[width=0.8\linewidth, center]{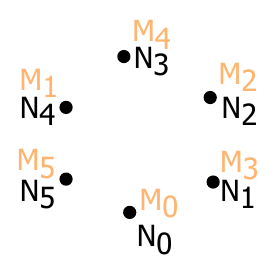}
  \caption{A possible DNN partition \\ placement chosen by network \\ orchestrator, optionally with\\
  app manager choosing the \\ slice fitness function. 
  }
  \label{fig:sfig4}
\end{subfigure}
\caption{Communication-aware model distribution.}
\label{fig:mdi}
\end{figure}

We next turn to a use case in inferencing that illustrates the architecture's support for application-network cooperation. 
More specifically, the application has a DNN model that it partitions into model parts. It derives an evaluation graph, $G$, over the parts that indicates the flow of information between the parts and their associated communication cost, cf.~Figure \ref{fig:sfig1}. The goal of the application manager is to achieve communication-aware distribution of the parts, so that the communication between the parts is not adversely affected by the available network capacity between the nodes hosting the parts \cite{bhardwaj2019memory, jian2023communication}.

Assume there is a network orchestrator,  published as the function $F_{Orchestrator}$, that can optimize the choice of network slice for a distributed model and deploy the model on that slice. The function's optimization can be based on various generic fitness criteria that the application can choose from. In this case, the application would choose bandwidth between the nodes and possibly latency as its fitness criteria. By publishing $G$ as named data and subscribing to the information specification $I_{MP}$
\begin{equation*}
    I_{MP} = (G, F_{Orchestrator}) \; ,
\end{equation*}
the application manager can delegate the task of deploying and launching the partitioned model, in a fashion that makes the network aware of the application communication needs, while also incorporating network aware slice selection.  Figure~\ref{fig:sfig4} shows a possible partition placement resulting from the subscription to $I_{MP}$.

A more sophisticated version of the orchestrator could permit the application to use third-party network vendor-published fitness functions, which could be used by the orchestrator for evaluating network slices.  This would be in line with the recent trend of disaggregating network functions, which is aimed at facilitating a marketplace of vendor-provided (and possibly intelligent) network support components. Assume there is a fitness function $F_{Fitness}$ published, for evaluating a particular type of fitness on a network slice, such as the one depicted in Figure~\ref{fig:sfig3}. For an orchestrator capable of leveraging such a function, the information subscription would become
\begin{equation*}
    I_{MP} = ((G, F_{Fitness}), F_{Orchestrator}) \, .
\end{equation*}

By enabling applications to specify their communication constraints as part of the information abstraction, the AI-EDGE architecture standardizes awareness sharing in a manner that is agnostic to network implementation. In other words, the information specifications published by the application remain unchanged across an any network that implements the AI-EDGE reference architecture. The underlying network's implementation of the $F_{Orchestrator}$ and $F_{Fitness}$ functions can vary but as long as their API contract remains unchanged, the information specifications remain valid.

\subsection{Security xApp for O-RAN 5G Network}

Our last use case considers 5G-cellular networking at the edge and explores leverage of the AI-EDGE architecture in the context of Open Radio Access Network (O-RAN). O-RAN, under the stewardship of the O-RAN Alliance, has emerged as a primary 5G architecture and seen widespread operator adoption worldwide \cite{polese2023understanding}. It incorporates RAN Intelligent Controllers (RICs) that provide network management and control at varying time scales---near real-time (nearRT) and non real-time (nonRT) \cite{polese2023understanding}. The RIC architecture specifies open and standard interfaces for applications in the RIC to subscribe to telemetry from RAN nodes and to transmit control signals down to the nodes. Applications in the nearRT RIC and nonRT RIC are referred to as xApps and rApps respectively. RIC is aimed at accelerating the development of intelligent network control applications that can be integrated into the network as plug-and-play xApps and rApps. In particular, in O-RAN, \textit{E2} is the interface between the nearRT RIC and RAN nodes (E2 nodes) ---distributed units (DUs) and centralized units (CUs)--- and the E2 Service Model (\textit{E2SM}) instruments telemetry subscription, reporting and control over this interface. In turn, each xApp can use a custom service model whose agents in the RIC and the RAN nodes can use or extend an existing E2SM to subscribe to the type and granularity of telemetry specific to the xApp. 

\begin{figure}[t]
\begin{center}
\includegraphics[width=0.475\textwidth]{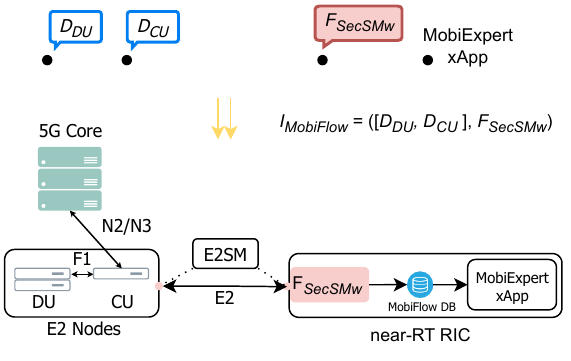}
\end{center}
\caption{Information-level specification of a security xApp in the reference architecture that captures data flow and function composition while hiding O-RAN RIC architecture's E2SM setup and subscription procedures, UE setup messages at E2 nodes producing the xApp telemetry, etc.}
\label{fig:ric-xapp-iw}
\end{figure}

Our use case explores \textit{MobiExpert} \cite{wen20245g}, a network security xApp that monitors control messages exchanged by user equipment (UE) and RAN nodes during initial setup to detect malicious UEs, fake base stations and man-in-the-middle attacks that threaten privacy of network users and disrupt network services. A prior implementation of the xApp directly in O-RAN has customized a bespoke service model ---let's call it \textit{SecSM}--- which filters for telemetry related to setup messages shared between UEs and RAN nodes, and generates records of these network flows called \textit{MobiFlow} records. \textit{MobiExpert} detects attacks using these records as input.

{\em MobiExpert} can alternatively be redesigned more simply by using an AI-EDGE architecture instantiated to support O-RAN platforms. The redesign uses an information specification, as follows.  
Assume $D_{DU}$ and $D_{CU}$ are named data published in the information waist corresponding to the telemetry available at each E2 node and $F_{SecSMw}$ is a named wrapper function in the RIC that instantiates the E2SM to subscribe and filter for the telemetry needed by {\em MobiExpert} and provides the xApp with the resulting telemetry flow records. The information required is then specified as:
    \begin{equation*}
        I_{MobiFlow} = ([D_{DU}, D_{CU}], F_{SecSMw}) 
    \end{equation*}

The wrapper function $F_{SecSMw}$ leverages a platform-specific translator that can be generically used by xApps other than {\em MobiExpert}. Developing and integrating this translator to configure the relevant agents on the RIC as well as the E2 Node is a one-time task, which is part of enabling support of the AI-EDGE architecture on an O-RAN Reference Platform. The resulting platform simplifies the development of xApps: the {\em MobiExpert} developer in this case does not have to understand the E2 interface and customize its E2SM service model. Using the information waist hides these details, providing a convenient abstraction for the developer integrating intelligence into O-RAN.

\section{Discussion, Future Work and Conclusions}
\label{sec:openqs}

We have presented an intelligence plane architecture based on core abstractions, which were developed in concert with partner stakeholders and illustrated through their use cases.  These abstractions are expected to yield a variety of tangible benefits for AI and network researchers, designers, developers, and operators, per the rationale we presented in Section~\ref{sec:rationale}. Arguably, these abstractions may also be beneficial for non-AI applications. However, we believe that AI uses are going to be the primary driver for the adoption of the abstractions, given the challenges and opportunities they are posing for today's internet.

In a sense, the information waist offers an alternative to data transports that are currently available in the internet. Dependability of the waist is thus a key consideration beyond its primary requirements of security and privacy. Implementations of the waist should therefore leverage standard principles of distributed algorithms to be inherently tolerant to the common failure modes of the network, including loss, reordering, and duplication of messages and crash, fail-stop, and byzantine behavior of nodes.  Likewise, provisioning for observability so that tools for fault detection and diagnostics can be used is of interest. In the long term, such observability can become a basis for data to relearn the functions in the intelligence plane.
    
The AI-EDGE architecture needs further study for heterogeneous and multi-operator networks: In the extreme case, for reasons of backwards compatibility, some networks may not support any of these abstractions. Other networks may, for reasons of security or privacy, choose to share no information from their intelligence plane with other networks or, for because of platform-specific implementations, have different APIs or semantics. Case studies evaluating inter-operator leverage of the architecture are needed.  Similarly, yet other networks may have intelligence plane implementations that share misleading information with other networks. Thus, a related topic for further study is how the intelligence plane of multiple such heterogeneous networks can interact and interoperate meaningfully. One approach for dealing with heterogeneity of networks is based on reasoning wherein the intelligence plane of a network can logically deduce the trustworthiness and usability of information shared by another network \cite{shieh2011netquery}. This could require disparate implementations of the intelligence plane to agree on a common mechanism of disseminating capabilities and supported services among the planes.

A final comment concerns heterogeneous data and models. For these to be meaningfully used in networks ---including the intelligence plane--- and applications, considerations of provenance and explainability are warranted to inform users' decisions. One option is to use metadata for addressing such considerations, just as we did for capturing the micro-architectural support for where models could be executed. Similarly, metadata about the format and resolution of data can inform their use within the intelligence plane.

\section*{Acknowledgment}
This work is partially supported by the National Science Foundation awards CNS-2112471, CNS-2130889, 
and OSU's 5G-OH Center. 

\bibliographystyle{IEEEtran}
\bibliography{aa-references, refs}

\begin{thebibliography}{10}
\providecommand{\url}[1]{#1}
\csname url@samestyle\endcsname
\providecommand{\newblock}{\relax}
\providecommand{\bibinfo}[2]{#2}
\providecommand{\BIBentrySTDinterwordspacing}{\spaceskip=0pt\relax}
\providecommand{\BIBentryALTinterwordstretchfactor}{4}
\providecommand{\BIBentryALTinterwordspacing}{\spaceskip=\fontdimen2\font plus
\BIBentryALTinterwordstretchfactor\fontdimen3\font minus
  \fontdimen4\font\relax}
\providecommand{\BIBforeignlanguage}[2]{{%
\expandafter\ifx\csname l@#1\endcsname\relax
\typeout{** WARNING: IEEEtran.bst: No hyphenation pattern has been}%
\typeout{** loaded for the language `#1'. Using the pattern for}%
\typeout{** the default language instead.}%
\else
\language=\csname l@#1\endcsname
\fi
#2}}
\providecommand{\BIBdecl}{\relax}
\BIBdecl

\bibitem{letaief2019roadmap}
K.~B. Letaief, W.~Chen, Y.~Shi, J.~Zhang, and Y.-J.~A. Zhang, ``The roadmap to
  {6G}: {AI} empowered wireless networks,'' \emph{IEEE Communications
  Magazine}, vol.~57, no.~8, pp. 84--90, 2019.

\bibitem{clark2003knowledge}
D.~D. Clark, C.~Partridge, J.~C. Ramming, and J.~T. Wroclawski, ``A knowledge
  plane for the internet,'' in \emph{Proceedings of the Conference on
  Applications, Technologies, Architectures, and Protocols for Computer
  Communications}, 2003, pp. 3--10.

\bibitem{bullot2008situatedness}
T.~Bullot, R.~Khatoun, L.~Hugues, D.~Ga{\"\i}ti, and L.~Merghem-Boulahia, ``A
  situatedness-based knowledge plane for autonomic networking,''
  \emph{International Journal of Network Management}, vol.~18, no.~2, pp.
  171--193, 2008.

\bibitem{manoj2008cognet}
B.~Manoj, R.~R. Rao, and M.~Zorzi, ``Cognet: a cognitive complete knowledge
  network system,'' \emph{IEEE Wireless Communications}, vol.~15, no.~6, pp.
  81--88, 2008.

\bibitem{camelo2022daemon}
M.~Camelo, M.~Gramaglia, P.~Soto, L.~Fuentes, J.~Ballesteros,
  A.~Bazco-Nogueras, G.~Garcia-Aviles, S.~Latr{\'e}, A.~Garcia-Saavedra, and
  M.~Fiore, ``Daemon: A network intelligence plane for {6G} networks,'' in
  \emph{2022 IEEE Globecom Workshops (GC Wkshps)}.\hskip 1em plus 0.5em minus
  0.4em\relax IEEE, 2022, pp. 1341--1346.

\bibitem{mestres2017knowledge}
A.~Mestres, A.~Rodriguez-Natal, J.~Carner, P.~Barlet-Ros, E.~Alarc{\'o}n,
  M.~Sol{\'e}, V.~Munt{\'e}s-Mulero, D.~Meyer, S.~Barkai, M.~J. Hibbett
  \emph{et~al.}, ``Knowledge-defined networking,'' \emph{ACM SIGCOMM Computer
  Communication Review}, vol.~47, no.~3, pp. 2--10, 2017.

\bibitem{ashtari2022knowledge}
S.~Ashtari, I.~Zhou, M.~Abolhasan, N.~Shariati, J.~Lipman, and W.~Ni,
  ``Knowledge-defined networking: Applications, challenges and future work,''
  \emph{Array}, vol.~14, p. 100136, 2022.

\bibitem{ghodsi2011information}
A.~Ghodsi, S.~Shenker, T.~Koponen, A.~Singla, B.~Raghavan, and J.~Wilcox,
  ``Information-centric networking: seeing the forest for the trees,'' in
  \emph{Proceedings of the 10th ACM Workshop on Hot Topics in Networks}, 2011,
  pp. 1--6.

\bibitem{ahlgren2012survey}
B.~Ahlgren, C.~Dannewitz, C.~Imbrenda, D.~Kutscher, and B.~Ohlman, ``A survey
  of information-centric networking,'' \emph{IEEE Communications Magazine},
  vol.~50, no.~7, pp. 26--36, 2012.

\bibitem{sifalakis2014information}
M.~Sifalakis, B.~Kohler, C.~Scherb, and C.~Tschudin, ``An information centric
  network for computing the distribution of computations,'' in
  \emph{Proceedings of the 1st ACM Conference on Information-Centric
  Networking}, 2014, pp. 137--146.

\bibitem{krol2019compute}
M.~Kr{\'o}l, S.~Mastorakis, D.~Oran, and D.~Kutscher, ``Compute first
  networking: Distributed computing meets icn,'' in \emph{Proceedings of the
  6th ACM Conference on Information-Centric Networking}, 2019, pp. 67--77.

\bibitem{mastorakis2020icedge}
S.~Mastorakis, A.~Mtibaa, J.~Lee, and S.~Misra, ``{IC}edge: When edge computing
  meets information-centric networking,'' \emph{IEEE Internet of Things
  Journal}, vol.~7, no.~5, pp. 4203--4217, 2020.

\bibitem{mtibaa2018towards}
A.~Mtibaa, R.~Tourani, S.~Misra, J.~Burke, and L.~Zhang, ``Towards edge
  computing over named data networking,'' in \emph{2018 IEEE International
  Conference on Edge Computing (EDGE)}.\hskip 1em plus 0.5em minus 0.4em\relax
  IEEE, 2018, pp. 117--120.

\bibitem{jacobson2009networking}
V.~Jacobson, D.~K. Smetters, J.~D. Thornton, M.~F. Plass, N.~H. Briggs, and
  R.~L. Braynard, ``Networking named content,'' in \emph{Proceedings of the 5th
  International Conference on Emerging Networking Experiments and
  Technologies}, 2009, pp. 1--12.

\bibitem{ghodsi2011naming}
A.~Ghodsi, T.~Koponen, J.~Rajahalme, P.~Sarolahti, and S.~Shenker, ``Naming in
  content-oriented architectures,'' in \emph{Proceedings of the ACM SIGCOMM
  Workshop on Information-Centric Networking}, 2011, pp. 1--6.

\bibitem{zhang2014named}
L.~Zhang, A.~Afanasyev, J.~Burke, V.~Jacobson, K.~Claffy, P.~Crowley,
  C.~Papadopoulos, L.~Wang, and B.~Zhang, ``Named data networking,'' \emph{ACM
  SIGCOMM Computer Communication Review}, vol.~44, no.~3, pp. 66--73, 2014.

\bibitem{tschudin2014named}
C.~Tschudin and M.~Sifalakis, ``Named functions and cached computations,'' in
  \emph{Proceedings of the 11th Consumer Communications and Networking
  Conference (CCNC)}.\hskip 1em plus 0.5em minus 0.4em\relax IEEE, 2014, pp.
  851--857.

\bibitem{ferguson2021orion}
A.~D. Ferguson, S.~Gribble, C.-Y. Hong, C.~Killian, W.~Mohsin, H.~Muehe,
  J.~Ong, L.~Poutievski, A.~Singh, L.~Vicisano \emph{et~al.}, ``Orion: Google's
  $\{$Software-Defined$\}$ networking control plane,'' in \emph{18th USENIX
  Symposium on Networked Systems Design and Implementation (NSDI 21)}, 2021,
  pp. 83--98.

\bibitem{alliance2020ran}
O.~{A}lliance, ``{O-RAN W}orking {G}roup 3: {N}ear-real-time {R}an intelligent
  controller-{E2} {S}ervice {M}odel ({E2SM}),'' \emph{{ORAN}-{WG3}.
  E2SM-KPM-v01}, 2020.

\bibitem{o2021ran}
S.~{O-RAN} {W}orking~{G}roup 1, ``{O-RAN} architecture description 5.00,''
  \emph{{O-RAN}. {WG1}. {O-RAN-A}rchitecture-{D}escription-v05. 00 Technical
  Specification}, 2021.

\bibitem{katsis2025zt}
C.~Katsis and E.~Bertino, ``{ZT-SDN}: an {ML}-powered zero-trust architecture
  for software-defined networks,'' \emph{ACM Transactions on Privacy and
  Security}, vol.~28, no.~2, pp. 1--35, 2025.

\bibitem{thompson2014ndn}
J.~Thompson and J.~Burke, ``{NDN} common client libraries,'' \emph{Technical
  Report NDN-0024, Revision 1. NDN Project}, 2014.

\bibitem{newberry2021yanfd}
E.~Newberry, X.~Ma, and L.~Zhang, ``{YaNFD}: yet another named data networking
  forwarding daemon,'' in \emph{Proceedings of the 8th ACM Conference on
  Information-Centric Networking}, 2021, pp. 30--41.

\bibitem{scherb2019execution}
C.~Scherb, C.~Marxer, and C.~Tschudin, ``Execution plans for serverless
  computing in information centric networking,'' in \emph{Proceedings of the
  1st ACM CoNEXT Workshop on Emerging in-Network Computing Paradigms}, 2019,
  pp. 34--40.

\bibitem{kumamoto2020real}
Y.~Kumamoto, H.~Yoshii, and H.~Nakazato, ``Real-world implementation of
  function chaining in named data networking for {IoT} environments,'' in
  \emph{2020 IEEE International Workshop Technical Committee on Communications
  Quality and Reliability (CQR)}.\hskip 1em plus 0.5em minus 0.4em\relax IEEE,
  2020, pp. 1--6.

\bibitem{nour2021access}
B.~Nour, H.~Khelifi, R.~Hussain, S.~Mastorakis, and H.~Moungla, ``Access
  control mechanisms in named data networks: A comprehensive survey,''
  \emph{Acm computing Surveys (cSuR)}, vol.~54, no.~3, pp. 1--35, 2021.

\bibitem{kekki2018mec}
S.~Kekki, W.~Featherstone, Y.~Fang, P.~Kuure, A.~Li, A.~Ranjan, D.~Purkayastha,
  F.~Jiangping, D.~Frydman, G.~Verin \emph{et~al.}, ``{MEC} in {5G} networks,''
  \emph{ETSI white paper}, vol.~28, no. 2018, pp. 1--28, 2018.

\bibitem{3gppedgeapp}
N.~Gupta, ``Study on application architecture for enabling edge applications,''
  in \emph{S6-190111, 3GPP TSG-SA WG6 Meeting}, vol.~28, pp. 21--25.

\bibitem{righteous}
A.~Rakshit, S.~Reddy, R.~Ramnath, A.~Arora, and J.~Boubin, ``Righteous:
  Automatic right-sizing for complex edge systems,'' in \emph{Proceedings of
  the 9th ACM/IEEE Symposium on Edge Computing}.\hskip 1em plus 0.5em minus
  0.4em\relax ACM, 2024.

\bibitem{zhang2021sinan}
Y.~Zhang, W.~Hua, Z.~Zhou, G.~E. Suh, and C.~Delimitrou, ``Sinan: {ML}-based
  and {QoS}-aware resource management for cloud microservices,'' in
  \emph{Proceedings of the 26th ACM International Conference on Architectural
  Support for Programming Languages and Operating Systems}, 2021, pp. 167--181.

\bibitem{shu2020novel}
Z.~Shu and T.~Taleb, ``A novel {QoS} framework for network slicing in {5G} and
  beyond networks based on {SDN} and {NFV},'' \emph{IEEE Network}, vol.~34,
  no.~3, pp. 256--263, 2020.

\bibitem{d2022orchestran}
S.~D’Oro, L.~Bonati, M.~Polese, and T.~Melodia, ``Orchest{RAN}: Network
  automation through orchestrated intelligence in the open {RAN},'' in
  \emph{IEEE INFOCOM 2022-IEEE Conference on Computer Communications}.\hskip
  1em plus 0.5em minus 0.4em\relax IEEE, 2022, pp. 270--279.

\bibitem{yu2021jointly}
L.~Yu, R.~Albelaihi, X.~Sun, N.~Ansari, and M.~Devetsikiotis, ``Jointly
  optimizing client selection and resource management in wireless federated
  learning for {I}nternet of {T}hings,'' \emph{IEEE Internet of Things
  Journal}, vol.~9, no.~6, pp. 4385--4395, 2021.

\bibitem{qu2022context}
Z.~Qu, R.~Duan, L.~Chen, J.~Xu, Z.~Lu, and Y.~Liu, ``Context-aware online
  client selection for hierarchical federated learning,'' \emph{IEEE
  Transactions on Parallel and Distributed Systems}, vol.~33, no.~12, pp.
  4353--4367, 2022.

\bibitem{deng2021fair}
Y.~Deng, F.~Lyu, J.~Ren, Y.-C. Chen, P.~Yang, Y.~Zhou, and Y.~Zhang, ``Fair:
  Quality-aware federated learning with precise user incentive and model
  aggregation,'' in \emph{IEEE INFOCOM 2021-IEEE Conference on Computer
  Communications}.\hskip 1em plus 0.5em minus 0.4em\relax IEEE, 2021, pp.
  1--10.

\bibitem{bhardwaj2019memory}
K.~Bhardwaj, C.-Y. Lin, A.~Sartor, and R.~Marculescu, ``Memory-and
  communication-aware model compression for distributed deep learning inference
  on {IoT},'' \emph{ACM Transactions on Embedded Computing Systems (TECS)},
  vol.~18, no.~5, pp. 1--22, 2019.

\bibitem{jian2023communication}
T.~Jian, D.~Roy, B.~Salehi, N.~Soltani, K.~Chowdhury, and S.~Ioannidis,
  ``Communication-aware {DNN} pruning,'' in \emph{IEEE INFOCOM 2023-IEEE
  Conference on Computer Communications}.\hskip 1em plus 0.5em minus
  0.4em\relax IEEE, 2023, pp. 1--10.

\bibitem{polese2023understanding}
M.~Polese, L.~Bonati, S.~D’oro, S.~Basagni, and T.~Melodia, ``Understanding
  {O-RAN}: Architecture, interfaces, algorithms, security, and research
  challenges,'' \emph{IEEE Communications Surveys \& Tutorials}, vol.~25,
  no.~2, pp. 1376--1411, 2023.

\bibitem{wen20245g}
H.~Wen, P.~Porras, V.~Yegneswaran, A.~Gehani, and Z.~Lin, ``{5G-SPECTOR}: an
  {O-RAN} compliant {Layer-3} cellular attack detection service,'' in
  \emph{Network and Distributed System Security (NDSS) Symposium}, 2024.

\bibitem{shieh2011netquery}
A.~Shieh, E.~G. Sirer, and F.~B. Schneider, ``{NetQuery}: A knowledge plane for
  reasoning about network properties,'' in \emph{Proceedings of the ACM SIGCOMM
  2011 Conference}, 2011, pp. 278--289.

\end{thebibliography}

\end{document}